\documentclass[twocolumn,showpacs,aps,pra,a4paper]{revtex4}
\usepackage{epsfig,amssymb,amsmath,amsmath}

\newcommand{\hc}{\hat{c}}
\newcommand{\hd}{\hat{d}}
\newcommand{\ra}{\rangle}
\newcommand{\la}{\langle}
\newcommand{\hN}{\hat{N}}

\date{\today}
\begin{document}


\title{Entangling atoms in bad cavities}

\author{Anders S\o ndberg S\o rensen and Klaus M\o lmer}
\affiliation{QUANTOP, Danish quantum optics center\\ Institute of
  Physics and Astronomy, University of Aarhus, 
DK-8000 Aarhus C, Denmark}

\begin{abstract}
We propose a method to produce entangled spin squeezed states
of a large number of atoms inside an optical cavity. By illuminating
the atoms 
with bichromatic light, the coupling to the cavity induces pairwise
exchange of excitations which entangles the atoms.
Unlike most proposals for
entangling atoms by cavity QED, our proposal does not require the
strong coupling regime $g^2/\kappa\Gamma\gg 1$, where $g$ is the atom
cavity coupling strength, $\kappa$ is the cavity decay rate, and
$\Gamma$ is the decay rate of the atoms. In this work the important
parameter is $N g^2 /\kappa\Gamma$, where $N$ is the number of atoms,
and our proposal permits the
production of entanglement in bad cavities as long as they contain a
large number 
of atoms. 
\end{abstract}


\maketitle

\section{Introduction}
\label{sec:intro}
To obtain a large coherent coupling of individual quantum
systems while at the same time maintaining a low decoherence rate
is the main challenge in
the experimental exploration of entanglement. An example of this is
cavity QED which was one of the first 
proposals for the 
construction of a quantum computer and creation of entanglement of
atoms \cite{pellizari,domokos}, but where the experimental progress has been
hampered by 
decoherence caused by cavity decay and spontaneous emission from the atoms.
To overcome these problems experiments have resorted to very small optical
cavities where the small cavity volume increases the interaction
strength \cite{kimble,rempe}, or Rydberg atoms in superconducting microwave 
cavities, where the decoherence rates are low \cite{haroche,walther}. 
 
Most cavity QED schemes which have been proposed so far require the
cavity to 
be in a strong coupling regime $g^2/\kappa\Gamma\gg 1$,
where $g$ is the atom
cavity coupling strength, $\kappa$ is the cavity decay rate, and
$\Gamma$ is the decay rate of the atoms, and this limit is very hard
to achieve experimentally.  In this paper we propose a
scheme where the created entanglement depends on the parameter
$Ng^2/\kappa\Gamma$, where $N$ 
is the number of atoms. With this scheme it is in principle possible to
produce entanglement in any cavity, but in practise the entanglement
becomes unmeasurable if $Ng^2/\kappa\Gamma\ll 1$. On the other hand, if
$Ng^2/\kappa\Gamma \gtrsim 1$ a measurable entanglement is
produced. Compared to the requirement  $g^2/\kappa\Gamma\gg 1$ the
present approach thus allows a substantially reduction in the requirements
for the cavity if a large number of atoms is used. We note that the
requirement $Ng^2/\kappa\Gamma \gtrsim 1$ is equivalent to the
criterion for optical bistability and squeezing in cavity QED as studied
experimentally in Refs. 
\cite{rempe91,raizen,grangier}.   

We propose to produce so-called spin
squeezed states \cite{ueda}. The collective properties of $N$ two
level atoms are conveniently described by pseudo angular momentum
operators defined by $J_z=\sum_k (|a\ra_k \la a|-|b\ra_k \la b|)/2$ and
$J_+=\sum_k |a\ra_k \la b|$, where the sum is over the individual
atoms, and where $|a\ra$ and $|b\ra$ are the two internal states of
the atoms. The state where all atoms
are in the $a$ state is an eigenstate of the $J_z$ operator with
eigenvalue $N/2$. If the $J_x$
operator is measured in this state, the result will fluctuate around
the mean value of zero with a variance $N/4$. By entangling the atoms
it is possible to maintain a large value of
the mean spin $\la J_z \ra\approx N/2$ while considerably reducing the
noise in 
a spin component $J_\theta=\cos(\theta)J_x+\sin(\theta)J_y$
perpendicular to the mean spin. A state with this property is
called a spin squeezed state. 

The experimental generation of spin
squeezed states has potential applications in high precision spectroscopy
and atomic clocks. 
For spectroscopy on a collection of two-level atoms 
Wineland {\it et al.} \cite{wineland} have shown that the possible
gain in precision by using a spin squeezed state is given by the quantity  
\begin{equation}
\xi^2=\underset{\theta}{{\rm min}}{\left( \frac{N\langle J_\theta^2
  \ra}{\langle J_z\rangle^ 2}\right)}.  
\label{eq:winxi}
\end{equation}
It has also been shown that $\xi^2<1$ indicates that the atom are in an
entangled state \cite{condensate,extreme}, and here we shall use
$\xi^2$ to characterize the entanglement of the atoms. 

Several schemes for the production of spin squeezed states
have been
proposed, and recently the first weakly
squeezed states have been produced experimentally by absorption of squeezed
light \cite{hald}, by QND-detection \cite{kuzmich}, and by collisional
interaction \cite{kasevich}. 
The possibility to produce spin squeezed states by having a large
number of atoms in a bad cavity was already proposed in Ref.\ 
\cite{vernac}, but the method proposed here is more
efficient; with the same cavity parameters our proposal enables a
stronger squeezing of the spin.  
During the preparation of this work we became aware that a scheme very
similar to ours has recently been proposed \cite{lukin}. Compared to
that work we apply a simpler level scheme 
in a different regime, but the fundamental ideas and the results are
very similar. Also the results reached in Ref.\ \cite{thomsen} are
similar to ours, but the mechanism employed in that paper is very
different from what we propose here. 

The paper is organized as follows. In Sec.\ \ref{sec:ideal} we present
our scheme in the ideal case where no dissipation takes place. In
Sec.\ \ref{sec:nonideal} we analyse the scheme in the more realistic
situation where cavity decay and decay of the excited atomic states
affect the preparation of the entangled state. In Sec.\
\ref{sec:conclusion} we summarize our proposal, and we discuss how
some states are more robust than others against dissipation and loss.
 The derivation of the
evolution in the presence of dissipation uses the standard method of
adiabatic elimination and produces quite complicated equations. These
expressions are not essential for the understanding of the functioning
of the proposal and we have put the technical derivation in the
Appendix.   

\section{Ideal case}
\label{sec:ideal}

\begin{figure}[b]
\begin{center}
 \begin{minipage}{4.1cm}
  \epsfig{file=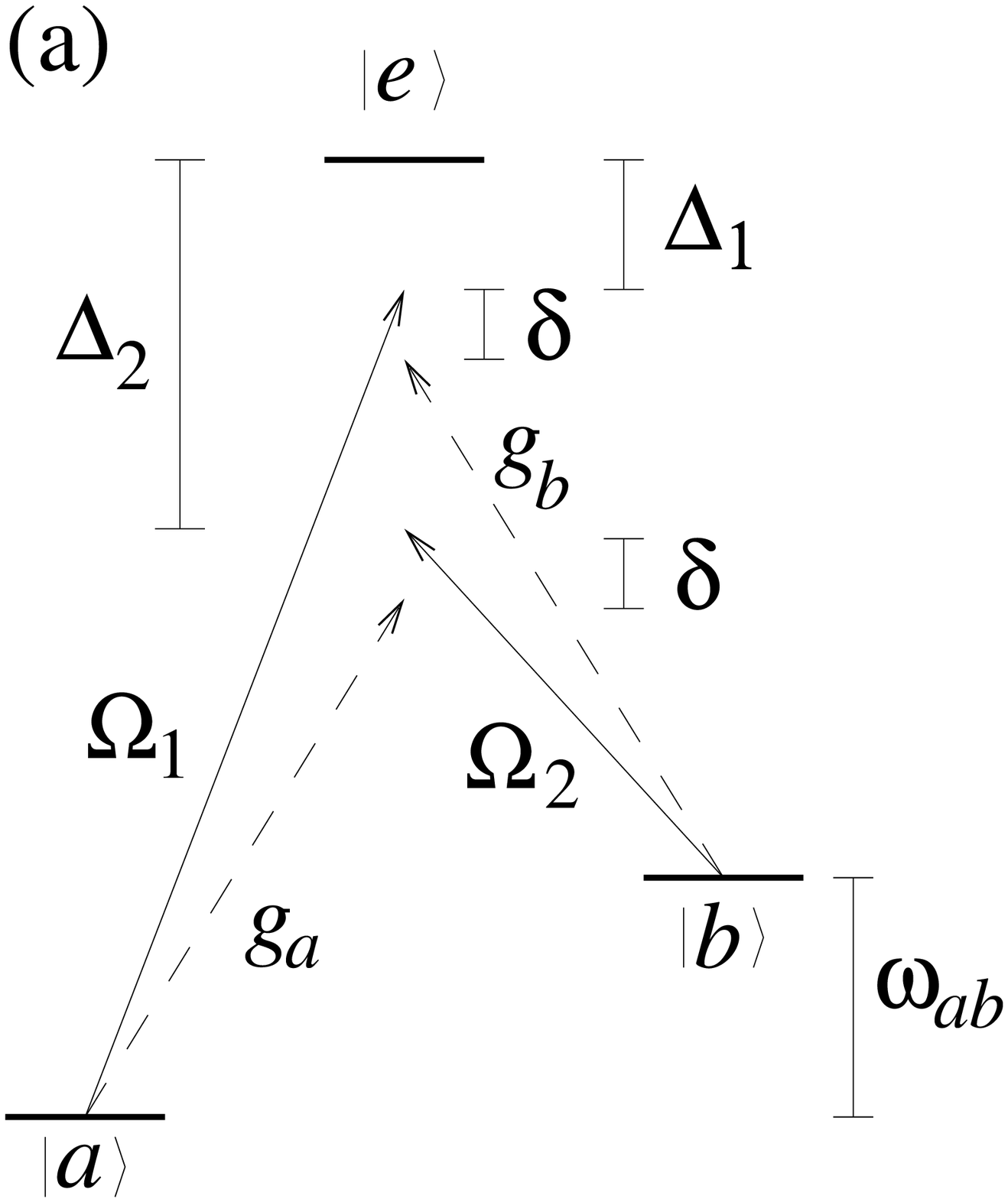,angle=0,width=4.1cm}
 \end{minipage}
\hspace{0.2cm}
 \begin{minipage}{4.1cm}
  \epsfig{file=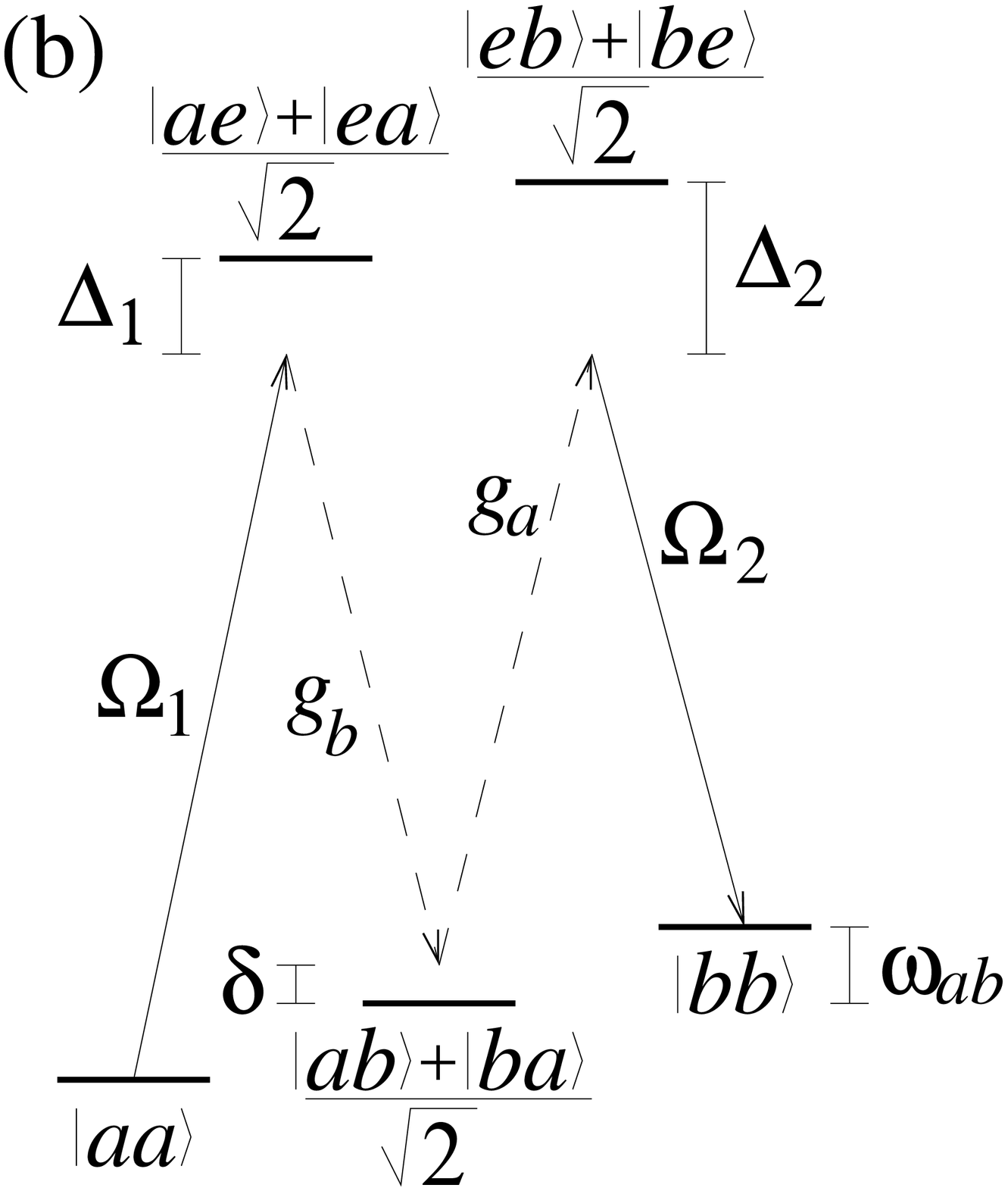,angle=0,width=4.1cm}
 \end{minipage}
\end{center}
\caption[]{Energy levels and couplings. (a) The energy levels of an
  atom and the couplings induced by the lasers and the cavity. 
  The laser and cavity couplings permits the pairwise transfer of atoms
  from state $|a\ra$ to $|b\ra$ as indicated in part (b). The dashed
  lines indicate transitions induced by the cavity and the full lines
  are the laser couplings.}
\label{fig:energy}
\end{figure}

The energy levels of the atoms and the laser couplings  are
depicted in Fig.\ \ref{fig:energy} (a).  
We consider a $\Lambda$ type three level atom with two stable
ground states $|a\rangle$ and $|b\rangle$ with an energy difference
$\omega_{ab}$ and an excited state $|e\ra$ with  energy difference
$\omega_{ae}$ to the
ground states $|a\ra$ ($\hbar=1$).
The state $|a\ra$ is coupled to the excited state $|e\ra$ by a laser
with a resonant Rabi frequency $\Omega_1$ and a frequency $\omega_1$
which is detuned from the excited state. Similarly the state $|b\ra$
is also coupled to the excited state by another detuned laser with
resonant Rabi frequency $\Omega_2$ and frequency $\omega_2$. The two
frequencies of the lasers are chosen such that their difference is 
exactly twice the energy splitting between the two
ground states $\omega_1-\omega_2=2\omega_{ab}$. With this choice of
frequencies all transitions involving only a single atom are
off-resonant, but a transition which transfers pairs of atoms from  
state $|a\ra$ to $|b\ra$ is resonant. A similar choice of resonance
conditions has also been proposed for trapped ions \cite{ion1,ion2},
and recently this scheme has allowed the first experimental production
of four particle entangled states \cite{sackett}.  To produce the
pairwise excitations of the atoms, it is not sufficient that the
process is resonant; it is also necessary that there exists a physical
mechanism which enables an interaction between the atoms. In
\cite{ion1,ion2} this was done by the Coulomb interaction between the
ions. Here we assume
that the quantized field in an optical cavity couples both the states 
$|a\ra$ and $|b\ra$ to the excited state $|e\ra$ with coupling
constants $g_a$ and $g_b$ respectively. With this coupling to the
cavity there exists a transition path for the pairwise transition, as
shown in Fig.\ \ref{fig:energy} (b), and the matrix element for the
transition becomes non-zero. In the remainder of this section we show
that this coupling leads to a spin squeezed state if we apply the
coupling to a state where all atoms are initially in the $a$ state. 

If we assume all fields to be propagating in the same direction, the
experimental  situation is described by the Hamiltonian 
\begin{equation}
\begin{split}
&H=\omega_0 \hc^\dagger \hc+ \sum_{k=1}^N \omega_{ae} |e\rangle_k
\langle e| +\omega_{ab} 
|b\rangle_k \langle b|+ H_{{\rm int},k} \\  
&
\begin{aligned}[t]
H_{{\rm int},k}=&{\left( \frac{\Omega_1} 2 {\rm e}^{-i\omega_1 t}+ g_a
    \hc
\right)} |e\rangle_k 
\langle a| 
\\& 
+{\left( \frac{\Omega_2} 2 {\rm e}^{-i\omega_2 t}+ g_b \hc
 \right)} |e\rangle_k
\langle b| + H.C. ,
\end{aligned}
\end{split}
\end{equation}
where $\hc$ and  $\omega_0$ denote the annihilation operator and frequency
of the relevant cavity mode. 

If we are in a regime where the laser power is sufficiently weak  that
we do not transfer any population to the excited atomic state we may
adiabatically eliminate this state and obtain an effective Hamiltonian
for the coupled state of the ground states and the cavity
\cite{james}. Assuming 
further that the lasers are also sufficiently weak that we do not create a
significant photon excitation in the cavity we may also adiabatically
eliminate the cavity field and we are left with an effective Hamiltonian for
the atoms \cite{james} 
\begin{equation}
\begin{split}
  H=\frac{1}{\delta}\bigglb(& \frac{|\Omega_1|^2|g_b|^2}{4 \Delta_1^2}
      J_+J_- + \frac{|\Omega_2|^2|g_a|^2}{4 \Delta_2^2}
      J_-J_+ \\ 
   &+ \frac{\Omega_1^* g_b g_a^* \Omega_2}{4 \Delta_1 \Delta_2}
      J_+J_+ + \frac{\Omega_2^* g_a g_b^* \Omega_1}{4 \Delta_1 \Delta_2}
      J_-J_-\biggrb),
\end{split}
\label{eq:Hideal}
\end{equation}
where we have omitted some unimportant energy shifts, and we have
introduced the detunings from the excited 
state $\Delta_1=\omega_{ae}-\omega_{1}$ 
and $\Delta_2=\omega_{ae}-\omega_{ab}- \omega_2=\Delta_1+\omega_{ab}$,
and the detuning from the cavity mode
$\delta=  \omega_1-\omega_{ab} -\omega_0= \omega_2+\omega_{ab}
-\omega_0$. These detunings are also defined in Fig.\
\ref{fig:energy}.
The angular momentum operators are defined as the angular momentum
operators in Sec.\ \ref{sec:intro}. 
The origin of each of the terms in this Hamiltonian can be understood
from processes like the one shown in Fig.\ \ref{fig:energy} (b) which
gives the term with $J_+J_+$, i.e., a double Raman process which takes
two atoms 
from $|a\ra$ to $|b\ra$ by absorption by $\Omega_1$, emission into the cavity
by $g_b$, reabsorption of the cavity photon by $g_a$, and emission by
$\Omega_2$.  

If we choose the strength of the  two Raman processes to  be identical
$\Omega_1 g_b^*/\Delta_1=\Omega_2 g_a^*/\Delta_2=\Omega g^*/\Delta$, Eq.\
(\ref{eq:Hideal}) reduces to the simpler form $H_{{\rm ideal}}=\chi
J_x^2$, where 
$\chi=|\Omega|^2|g|^2/\Delta^2\delta$. The squeezing arising from this
Hamiltonian can be calculated analytically \cite{ueda}. Starting from
an initial state where all atoms are in the $a$ state and propagating
with this Hamiltonian, squeezing by
a factor of 
$\xi^2\approx N^{-2/3}$ is produced (in the limit
$N\gg1$), and this is a significant noise reduction if a large
number of atoms is present. In the following section we show that a
significant squeezing is produced even in the presence of dissipation. 

The time it takes to produce a spin squeezed state of many atoms is very
short. 
To squeeze the spin by a constant factor, a constant
number of atoms has to be transfered into the $b$ state. With
increasing $N$ a decreasing fraction of the atoms has to be transfered,  and
thus a shorter 
time (scaling as $1/N$) is necessary to make the squeezing. 
The different decoherence mechanisms therefore have less time
to affect the preparation of the squeezed states, and as we show below,
this reduces the experimental requirement for the production of
squeezed states.

\section{Analysis including dissipation and noise}
\label{sec:nonideal}

The main purpose of this paper is to demonstrate that it is possible
to use a cavity to entangle atoms even in situations where 
substantial dissipation is present. In this section we analyse the
performance of our proposal in the presence of the two main decoherence
mechanisms: spontaneous emission and cavity
decay.
 
Before making a quantitative analysis of the effect of dissipation we
first make a few simple estimates of the decoherence.  For
simplicity we shall here assume that the lasers have 
approximately the same Rabi frequencies ($\Omega_l\sim\Omega$) and
detunings  ($\Delta_l\sim \Delta$) and also that the cavity couplings are 
similar $g_l\sim g$. The
number of spontaneously emitted photons is estimated to be approximately
$N_\Gamma\sim N \Gamma t |\Omega|^2 /\Delta^2$, where $\Gamma$ is the total
decay rate. The time required to produce  squeezing by
a constant factor is given by $t\sim \Delta^2\delta/(Ng^2\Omega^2)$,
and by inserting this expression we
find that the total number of decayed atoms is 
\begin{equation}
 N_\Gamma\sim \frac{\Gamma\delta}{g^2}.
\label{eq:ngamma}
\end{equation}
The number of photons decaying out of the cavity during the same time is
estimated to be $N_\kappa\sim N \kappa t \Omega^2g^2/(\Delta^2\delta^2)$
which reduces to
\begin{equation}
N_\kappa\sim \frac{\kappa}{\delta}
\label{eq:nkappa}
\end{equation}
when the expression for the time is inserted.

Because the spin squeezed state are only weakly entangled  they are
quite insensitive to the spontaneous emission of the atoms. If we
assume that we are near the initial state where $J_z\approx N/2$,
$J_x$ and $J_y$ may be replaced by the canonical conjugate position
$x=J_x\sqrt{2/N}$ and 
momentum $p=J_y\sqrt{2/N}$ operators of a harmonic oscillator. The
relaxation rate for the harmonic oscillator is then the same as
the relaxation rate for a single atom, and from the  well known
properties of squeezing of harmonic oscillators we find that the
squeezing is not completely degraded as long as the number of
decayed atoms is much less than the total number of atoms. The decay
of photons out of the cavity is more severe than the decay of a single
atom. Because the cavity couples to a collective degree of freedom,
the decoherence of the cavity will also affect the collective degree
of freedom. If $|\Omega_1 g_b|/\Delta_1=|\Omega_2g_a|/\Delta_2$ we
estimate that the first decay of a photon out of the cavity increases
the variance in all directions perpendicular to the mean spin by a
factor of three. To obtain a large 
squeezing we therefore require that at most a few photons are
scattered out of the cavity. From the expression in Eqs.\
(\ref{eq:ngamma}) and (\ref{eq:nkappa}) we see that we can fulfill
both $N_\Gamma\ll N$ and $N_\kappa\lesssim 1$ if the cavity
parameters fulfill
$N g^2 \gg \kappa\Gamma$ and we thus expect to be able to produce
substantial squeezing in this regime. This is confirmed by our more
accurate treatment of dissipation below.

Dissipation is described by the master equation for the density
matrix $\rho$ 
\begin{equation}
\frac{d}{dt} \rho = -i[H,\rho]+\frac{1}{2}\sum_{m} {\left(
    2\hd_m \rho \hd_m^\dagger
    -\hd_m^\dagger \hd_m \rho-\rho\hd_m^\dagger \hd_m \right)},  
\end{equation}
where $\hd_m$ are relaxation operators.
We assume that the separation of the atoms in the cavity is much
larger than the wavelength of the spontaneously emitted photons. In
this limit the decay of the atoms is uncorrelated and can be
described by independent relaxation operators for each atom. The
excited state $|e\ra$ is assumed to have three independent decay
channels: it may decay to the two lower states in the $\Lambda$-system
$|a\ra$ and $|b\ra$ with  decay rates
$\gamma_a$ and $\gamma_b$ respectively, and it may decay to some other
state  $|o\ra$ with a decay rate  $\gamma_o$. The total
effect of the spontaneous emission is described by $3N$ relaxation
operators
\begin{equation}
\begin{split}
\hd_{a,k}&=\sqrt{\gamma_a} |a\rangle_k\langle e|\\
\hd_{b,k}&=\sqrt{\gamma_b} |b\rangle_k\langle e|\\
\hd_{o,k}&=\sqrt{\gamma_o} |o\rangle_k\langle e|,
\end{split}
\label{eq:spontan}
\end{equation}
where $k=1,...,N$. To describe the decay of the cavity with a rate
$\kappa$ we introduce a relaxation operator
\begin{equation}
\hd_c=\sqrt{\kappa} \hc.
\label{eq:cavitydecay}
\end{equation}

To derive the evolution of the spin squeezing we first adiabatically
eliminate the excited state assuming 
\begin{equation}
\begin{split}
\frac{|\Omega_l|^2}{4}& \ll \Delta_l^2+\frac{\Gamma^2}4 \\
\delta, \kappa' &\ll \omega_{ab}
\end{split}
\label{eq:adiabatex}
\end{equation}
where $l=1,2$, and $\Gamma$ is the total decay rate  of the
excited state $\Gamma=\gamma_a+\gamma_b+\gamma_c$. $\kappa'$
is an effective decay rate of the cavity, which is 
slightly larger than $\kappa$ due to the scattering of cavity photons
by the atoms, cf.\ Eq.\ (\ref{eq:kappaprime}).  We then
adiabatically eliminate the cavity from the equations assuming 
\begin{equation}
N \frac{|\Omega_1g_b|^2}{4}\ll  {\left(
    \Delta_1^2+\frac{\Gamma^2}4\right)}  {\left(
    \delta^2+\frac{{\kappa'}^2}4\right)}.
\label{eq:adiabatcav}
\end{equation}

The equations resulting from the adiabatic elimination are quite
complicated and we leave the derivation to Appendix
\ref{sec:derivation}.
If we assume that the initial
state is almost unaffected by the interaction so that $J_z\approx N/2$,
the matrix elements which are quadratic in the angular momentum operators
do not couple to higher order terms and we obtain closed
equations for the expectation values
\begin{equation}
\frac{d}{dt}{\left[\begin{array}{c}
\langle J_z \rangle\\
\langle \hN_a+\hN_b\rangle\\
\langle J_+J_+\rangle\\
\langle J_-J_-\rangle\\
\langle J_+J_-\rangle\\
\langle J_-J_+\rangle
\end{array}\right]}=M\cdot
{\left[\begin{array}{c}
\langle J_z \rangle\\
\langle \hN_a+\hN_b\rangle\\
\langle J_+J_+\rangle\\
\langle J_-J_-\rangle\\
\langle J_+J_-\rangle\\
\langle J_-J_+\rangle
\end{array}\right]},
\label{eq:vectoreq}
\end{equation}
where $\hN_l$ is the number operator for atoms of type $l=a,b$
($\la \hN_a+\hN_b\ra$ is not a conserved quantity because of the decay to
the state $|o\ra$). The
precise form of the matrix $M$ is given by the 
expressions  in Eqs.\
(\ref{eq:Jz}-\ref{eq:A}).

\begin{figure}[bt]
\begin{center}
\epsfig{file=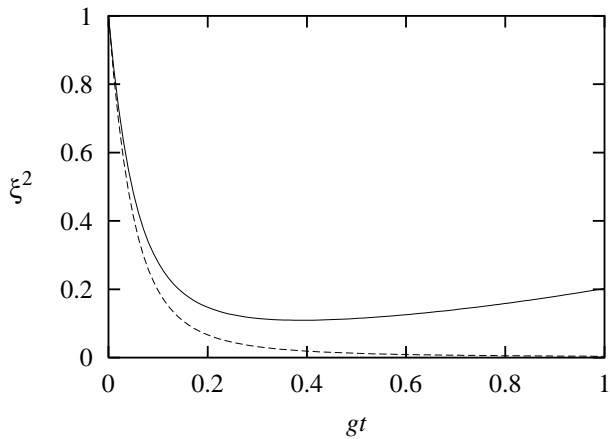,angle=270,width=8cm}
\end{center}
\caption[]{Evolution of squeezing for $N=10^6$ atoms in a bad
  cavity (full line). The parameters used in the simulation are
  $g_a=g_b=g$, 
  $\Gamma=\kappa=100  g$,
  $\gamma_a=\gamma_b=\gamma_o$, 
  $\Omega_1=\Omega_2=10^4 g$,
  $\Delta_1=10^5  g$, $\delta=5\cdot 10^2 g$, and  
  $\omega_{ab}=10^4 g$ corresponding to $g^2/\kappa\Gamma=10^{-4}$
  but $Ng^2/\kappa\Gamma=10^{2}$. For comparison we also show the
  evolution with 
  the same parameters but without dissipation, $\Gamma=\kappa=0$
  (dashed line).} 
\label{fig:evo}
\end{figure}

Due to the complicated structure of the matrix $M$ it is difficult to
describe the evolution of the squeezing analytically, but it is
straightforward to find the evolution numerically, where the solution
at a given time $t$ may by found by taking the exponential of the matrix $M
t$. In Fig.\ \ref{fig:evo} we show the evolution of squeezing in a
situation where $g^2/\Gamma\kappa\ll 1$ but  $N g^2/\Gamma\kappa\gg
1$. With the realistic parameters used in the figure we are able to
produce squeezing by approximately an order of magnitude after a very
short interaction time. The time required to make the squeezing in
the figure is less than  $1 \mu$s if we chose a realistic cavity
coupling parameter $g=(2\pi) 100$kHz.  

\begin{figure}[b]
  \centering
  \epsfig{file=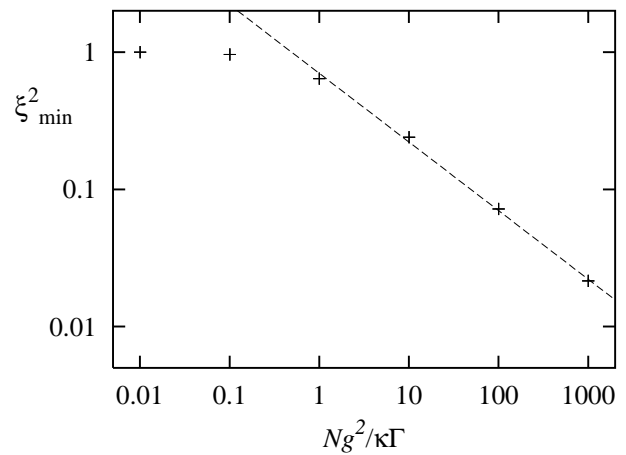,angle=270,width=8cm}
  \caption[]{Minimum squeezing parameter obtained by a numerical
    optimization. The points (+) are the results of the minimalization
    and the dashed line $0.7/\sqrt{Ng^2/\kappa\Gamma}$ 
    approximates $\xi_{{\rm min}}^2$ for  $Ng^2/\kappa\Gamma\gtrsim 1$.
    In the calculation we have
    assumed $N=10^6$, $g_a=g_b=g$,
    $\gamma_a=\gamma_b=\gamma_o$, and $\omega_{ab}=10^5$. The same
    minimum is obtained  for three different
    ratios $\kappa/\Gamma=10^{-2},1,$ and $10^2$, and the results are 
    independent of $N$ in the limit $N\gg 1$.}
  \label{fig:min}
\end{figure}

In principle the proposed scheme can be used to produce entanglement
in any cavity. The field $\Omega_2$ only couples to the ground state
$|b\ra$ that is initially unpopulated, so that a large value of
$\Omega_2$ increases the rate of the coherent transfer of atoms from
state $a$ to $b$ but does not affect the initial decoherence rate. 
The amount of squeezing, however,  depends
on the cavity parameters. To investigate the obtainable squeezing we
have performed a  numerical optimization of the
coupling strengths and detunings for a number of different cavity
parameters. With fixed values of the dissipation rates and energy
difference $\omega_{ab}$, we vary $\Omega_2/\Omega_1$, $\delta$, and
$\Delta_1$ and search for the values which give the minimal
$\xi^2$  (because all terms in $M$ involve the square of the
field strength, the minimum only depends on the ratio between the two
fields). In the limit $N\gg 1$ and for fixed ratios between the decay
rates $\gamma_l$ ($l=a,b,o$) the results of the optimization indicate
that the optimal squeezing parameter $\xi^2_{{\rm min}}$ is only a function
of the parameter 
$Ng^2/\Gamma \kappa$. In Fig.\ \ref{fig:min} we show $\xi^2_{{\rm min}}$ for
different cavity parameters. As expected from our simple estimates, the
figure confirms that strong squeezing can be produced in the limit
$Ng^2/\Gamma \kappa\gg 1$. In the calculations we have assumed
$g_a=g_b=g$ and 
$\gamma_a=\gamma_b=\gamma_o=\Gamma/3$, and with these values the
optimal squeezing is approximately given by
$\xi_{{\rm min}}^2=0.7/\sqrt{Ng^2/\Gamma \kappa}$ for $Ng^2/\Gamma
\kappa\gtrsim 1$. This is indicated by the dashed line in the figure. The
same behaviour but with a slightly different 
constant has also been found in \cite{lukin}. The obtained results only
change slightly if we vary the ratio between the coupling constants
or between the decay rates.   


The results of the numerical simulations agree very well with
the behaviour expected from the simple estimates. Our assumptions
of weak excitation of the atoms and of the cavity field mode
imply the necessary condition $\Delta_l \gg \Gamma$. 
It turns out, however,
that the detuning $\delta$ from the intermediate state in Fig.1 (b) 
with one cavity photon excited, does not need to be large.
If only the coupling is weak enough $\sqrt{N}|\Omega_1g_b/\Delta_1| \ll
\kappa$, and if the process that absorbs cavity photons is stronger than
the one producing them, $|\Omega_2g_a|/(\Delta_2^2+\Gamma^2/4)
> |\Omega_1g_b|/(\Delta_1^2+\Gamma^2/4)$, the photon excited 
state can be eliminated, and we find good squeezing for all values
of $\delta$. The minimum value of $\xi^2$ is found for $\delta=0$.

Finally, lets us briefly comment on a few experimental aspect of our
proposal.
The analysis above shows that our proposal is robust against the
spontaneous emission caused by the laser coupling and the decay of
photons out of the cavity. 
The coupling of the lasers, however, 
introduces other possible decoherence mechanisms if we are not able to
control the lasers with a high enough accuracy. In the treatment so
far we have ignored the AC-Stark shifts caused by the lasers because
they can be compensated by a small change in the frequencies. But the
magnitude 
of the AC-Stark depends on the power of the lasers so that fluctuations
in the power has a detrimental effect on the squeezing. To suppress this
effect we propose to adjust the relative strengths of the two fields so
that $|\Omega_1|^2\Delta_1/(\Delta_1^2+\Gamma^2/4)=
|\Omega_2|^2\Delta_2/(\Delta_2^2+\Gamma^2/4)$. With this choice the
AC-Stark shifts of the two ground states become identical and have no
effect on the internal state preparation. The problem of stabilizing
the power can thus be reduced to the problem of stabilizing the
relative frequency and intensity of the two fields which is much
easier experimentally if the two fields are derived from the same
source. 
 
The  efforts to entangle atoms through
cavity QED have so far concentrated on the strong coupling regime
$g^2/\kappa\Gamma\gg 1$. To achieve this limit it has been desirable
to use very small standing wave cavities where the coupling constant
varies sinusoidally along the cavity axis with a period of half the
optical wavelength. A controlled evolution in these cavities therefore
requires that the atoms  are localized in  regions smaller than the
optical wavelength. Since our proposal puts much less stringent
requirement on the cavity parameters it should not be necessary to use
such small cavities and  it could for instance be implemented with a
ring cavity. Then, the magnitude of the coupling
constant does not depend on the position of the atom along the cavity
axis, and if the classical fields are co-propagating with the cavity field
it is no longer necessary to localize the atoms within a wavelength. The atoms
only need to be confined within the waist of the cavity mode, and this
can be done with cold atoms trapped in a far detuned  optical dipole
trap or optical lattice, or even with atoms in a glass cell at room
temperature.  
  
\section{conclusion}
\label{sec:conclusion}
We have shown that it is possible to observe significant
spin squeezing of atoms coupled to the field mode in a 
lossy cavity. The loss of quantum correlations between the particles
which is caused by atomic decay and cavity loss is balanced
by the strong non-linear coupling achievable in the limit
of very many atoms. It has been shown that spin squeezing implies
entanglement, i.e., a separable state cannot lead to values
of $\xi^2$ smaller than unity. We have thus created entangled states
which are fairly robust against dissipation and loss. 

There is
no precise quantitative measure for the entanglement of a large 
collection of particles, but a natural qualitative measure is to
consider the possible gain, e.g., in spectroscopic resolution, that
the entangled states offers with respect to a disentangled state. By
binding the $N$ atoms together in $N/P$  maximally entangled
$P$-particle states $(|aaa ... a\rangle +|bbb ... b\rangle)/\sqrt{2}$,
one obtains a spectroscopic resolution corresponding to a state with
$\xi^2=1/P$ \cite{bollinger}. Hence in terms of spectroscopic
resolution the spin 
squeezed states are as powerful as if the atoms had been divided into
groups of maximally entangled
states of $1/\xi^2$ particles. 

From a practical perspective, however, the 
squeezed states offer a significant advantage compared to a
collection of highly entangled states. In  spin squeezed states 
the relevant observables are collective operators involving all the
atoms. There is
no need to address the atoms individually, and the manipulation and
detection of the squeezing can therefore 
be achieved by lasers addressing all atoms collectively. Furthermore the spin
squeezed states are
also easier to produce:
In an ideal spin squeezed state 
the one-particle density matrix $\rho_1$ is very close to the
initial pure state projection operator $\rho_1=|a\rangle \langle
a|+O(1/N)$, and the state of each atom is thus 
almost disentangled from the state of the other atoms. 
This means  that
we only need to perturb the initial state
slightly to turn it into a squeezed state. This is
a significant advantage in any experimental attempt to produce entanglement
because (a) the states are more robust against decoherence than
more highly entangled states and (b) the
interaction time 
required to make the desired state is much shorter than for the highly
entangled states.
The large number of atoms increases the decoherence
rates, i.e., more photons are scattered, but our calculations show
that the two advantages (a) and (b)  
outweigh the increased decoherence rate and enable the construction
of entangled states in situations where the experimental capabilities
do not permit the construction of entangled states of a few atoms.

In a broader context our proposal fits into the field of ensemble
quantum information processing, where the quantum information is
encoded into the collective degrees of freedom of a collection of
atoms. A number of papers \cite{duanprl,lukinensemble,duannature} have 
proposed schemes for the processing of information encoded
in such a way and the first experimental 
implementation of these concept has recently been reported
\cite{brian}. By combining these ideas with the present work 
one may for example
imagine a quantum computer with several separate and individually
addressable atomic clouds which communicate in a controlled manner
via cavity modes.

\begin{acknowledgments}
We are grateful to Michael Drewsen and Eugene Polzik for useful
discussions about possible experimental realizations of the scheme. 
This work was supported by the Danish National Research Foundation
through QUANTOP, the Danish Quantum Optics Center, and by 
CAUAC (contract no. HPRN-CT-2000-00165). 
\end{acknowledgments}

\appendix
\section{Deriving the equations of motion}
\label{sec:derivation}

In this section we derive the equations describing the time evolution
of squeezing in the presence of dissipation. We first consider only
the Hamiltonian describing a single atom and we adiabatically
eliminate the excited state of the atom by assuming that the population of
that state is negligible.
In this approximation the 
equations for the ground state density matrix elements are equivalent to
the evolution by a Hamiltonian 
\begin{equation}
\begin{split}
H=&-{\left(\frac{\Delta_1
       |\Omega_1|^2}{4{\left(\Delta_1^2+\frac{\Gamma^2}{4}\right)}}
     + \frac{\Delta_2
      |g_a|^2\hc^\dagger\hc}{\Delta_2^2+\frac{\Gamma^2}{4}}
     \right)}|a\ra \langle a|\\
&   -{\left(\frac{\Delta_2
       |\Omega_2|^2}{4{\left(\Delta_2^2+\frac{\Gamma^2}{4}\right)}}
     + \frac{\Delta_1
      |g_b|^2\hc^\dagger\hc }{\Delta_1^2+\frac{\Gamma^2}{4}}
     \right)}|b\ra \langle b|\\
& -\frac{\Delta_1}{\Delta_1^2+\frac{\Gamma^2}4}{\left(
     \frac{\Omega_1g_b^*}{2} |b\ra \langle a|  \hc^\dagger
   {\rm e}^{-i\delta t}+\frac{\Omega_1^*g_b}{2} |a\ra
   \langle b|  \hc 
   {\rm e}^{i\delta t}\right)}\\
& -\frac{\Delta_2}{\Delta_2^2+\frac{\Gamma^2}4}{\left(
     \frac{\Omega_2^*g_a}{2} |b\ra \langle a|  \hc
   {\rm e}^{i\delta t}+\frac{\Omega_2g_a^*}{2} |a\ra
   \langle b|  \hc^\dagger 
   {\rm e}^{-i\delta t}\right)}
\end{split}
\label{eq:Hgs}
\end{equation}
and six relaxation operators $\hd_{k,l}$ ($k=a,b,o$ and $l=1,2$) describing 
the combined excitation with a detuning $\Delta_l$ and decay to a
state $|k\ra$, e.g., 
\begin{equation}
\hd_{a,1}=\frac{\sqrt{\gamma_a}}{\Delta_1-i\frac{\Gamma}2}
{\left(\frac{\Omega_1}{2}|a\ra\la a|+g_b |a\ra\la b| \hc {\rm
      e}^{i\delta t} \right)}.
\end{equation}
To derive these result we have assumed that $\delta\ll \Delta_1,
\Delta_2$, and we have used the second relation in Eq.\
(\ref{eq:adiabatex}) to neglect processes which creates photons
without a change in the atomic state.

The first two lines in Eq.\ (\ref{eq:Hgs}) represent AC-Stark shifts
of the ground states. The first part of the shifts containing the
classical fields $\Omega_1$ and $\Omega_2$  can be compensated if we
make a change in the frequency of the fields. The second part
containing the quantum field $\hc$ is much smaller than the first and
by inserting the approximate time and Eq.\ (\ref{eq:hc}) below, we find
that this term gives a 
negligible phase shift if $g^2/\delta\Delta\ll 1$ and we shall
neglect these terms.  

We then adiabatically eliminate the cavity in the Heisenberg
picture. Setting $d(\hc {\rm e}^{i\delta t})/dt=0$ we obtain
\begin{equation}
\hc {\rm e}^{i\delta t}=-\frac{1}{\delta+i\frac{\kappa'}2}{\left( \frac
    {\Omega_1 g_b^*}{\Delta_1-i\frac{\Gamma}2} J_- +  \frac
    {\Omega_2 g_a^*}{\Delta_2-i\frac{\Gamma}2} J_+\right)}+{\rm noise},
\label{eq:hc}
\end{equation}
where the noise ensures the commutation relation of the operator. Here
we have introduced an effective decay rate for the cavity
\begin{equation}
\kappa'=\kappa+\frac{N\Gamma |g_a|^2}{\Delta_2^2+\frac{\Gamma^2}4}
\label{eq:kappaprime}
\end{equation}
which takes into account that the cavity photons may be scattered by
the atoms. In Eq.\ (\ref{eq:kappaprime}) we have assumed that
essentially all atoms remain in 
the $a$ state. 

The adiabatic elimination of the cavity requires that the atoms are
disentangled from the cavity, i.e., that
$\la \hc^\dagger \hc\ra \ll 1$. At $t=0$ this reduces to Eq.\
(\ref{eq:adiabatcav}), and at later times our numerical simulation
indicate that $\la \hc^\dagger \hc\ra$ typically changes slowly, so
that the condition is fulfilled if it is fulfilled at $t=0$.

Finally, we insert Eq.\ (\ref{eq:hc}) into the time derivatives of the
angular momentum operators, and by assuming that the initial state 
only changes slightly so that $J_z\approx N/2$, we obtain the following
expressions  
\begin{widetext}
\begin{equation}
\begin{split}
\frac{d}{dt}\langle J_z\rangle =&-
\frac{\gamma_b+\gamma_0/2}{\Delta_1^2+\frac{\Gamma^2}4}
\frac{|\Omega_1|^2}{4}\langle \hN_a \rangle
+\frac{\gamma_a+\gamma_0/2}{\Delta_2^2+\frac{\Gamma^2}4}
\frac{|\Omega_2|^2}{4}\langle \hN_b \rangle  \\
&-\frac{1}{\delta^2+\frac{{\kappa'}^2}4}\bigglb[
\begin{aligned}[t]
&\frac{|\Omega_1|^2|g_b|^2}{4(\Delta_1^2+\frac{\Gamma^2}4)^2}
{\left(-\delta\Delta_1(2\gamma_b+\gamma_o) +\kappa'\Delta_1^2+
\frac{\kappa'\Gamma(\gamma_a-\gamma_b)}4\right)} \langle J_+ J_- \rangle\\
&+\frac{|\Omega_2|^2|g_a|^2}{4(\Delta_2^2+\frac{\Gamma^2}4)^2}
{\left(\delta\Delta_2(2\gamma_a+\gamma_o) -\kappa'\Delta_2^2+
\frac{\kappa'\Gamma(\gamma_a-\gamma_b)}4\right)} \langle J_- J_+ \rangle\\
&+\frac{\Omega_1\Omega_2^*g_ag_b^*}{4(\Delta_2^2+\frac{\Gamma^2}{4})
  (\Delta_1^2+\frac{\Gamma^2}{4})} 
\begin{aligned}[t]
\bigglb(&-2i\delta\Delta_1\Delta_2
 -i\frac{\kappa'\Delta_1(\gamma_a+\frac{\gamma_o}2)}2 
 -i\frac{\kappa'\Delta_2(\gamma_b+\frac{\gamma_o}2)}2
 \\& 
 -\delta\Delta_1{\left(\gamma_b+\frac{\gamma_o}2\right)}
 +\delta\Delta_2{\left(\gamma_a+\frac{\gamma_o}2\right)}
+\frac{\kappa'\Gamma(\gamma_a-\gamma_b)}4\biggrb) \langle J_- J_- \rangle
\end{aligned}\\
&+\frac{\Omega_1^*\Omega_2g_a^*g_b}{4(\Delta_2^2+\frac{\Gamma^2}{4})
  (\Delta_1^2+\frac{\Gamma^2}{4})} 
\begin{aligned}[t]
\bigglb(&2i\delta\Delta_1\Delta_2+
 i\frac{\kappa'\Delta_1(\gamma_a+\frac{\gamma_o}2)}2  
 +i\frac{\kappa'\Delta_2(\gamma_b+\frac{\gamma_o}2)}2
\\&
  -\delta\Delta_1{\left(\gamma_b+\frac{\gamma_o}2\right)}
  +\delta\Delta_2{\left(\gamma_a+\frac{\gamma_o}2\right)}
+\frac{\kappa'\Gamma(\gamma_a-\gamma_b)}4\biggrb) \langle J_+ J_+
\rangle\biggrb] ,
\end{aligned}
\end{aligned}
\end{split}
\label{eq:Jz}
\end{equation}
%
%
\begin{equation}
\begin{split}
\frac{d}{dt} (\hN_a+\hN_b)=&-
\frac{\gamma_0}{\Delta_1^2+\frac{\Gamma^2}4}
\frac{|\Omega_1|^2}{4}\langle \hN_a \rangle
-\frac{\gamma_0}{\Delta_2^2+\frac{\Gamma^2}4}
\frac{|\Omega_2|^2}{4}\langle \hN_b \rangle  \\
&+\frac{\gamma_o}{\delta^2+\frac{{\kappa'}^2}4}\bigglb[
\begin{aligned}[t]
&\frac{|\Omega_1|^2|g_b|^2}{4(\Delta_1^2+\frac{\Gamma^2}4)^2}
{\left(2\delta\Delta_1 +\frac{\kappa'\Gamma}2\right)} \langle J_+ J_-
\rangle
+\frac{|\Omega_2|^2|g_a|^2}{4(\Delta_2^2+\frac{\Gamma^2}4)^2}
{\left(2\delta\Delta_2 +\frac{\kappa'\Gamma}2  \right)} \langle J_-
J_+ \rangle\\ 
&+\frac{\Omega_1\Omega_2^*g_ag_b^*}{4(\Delta_2^2+\frac{\Gamma^2}{4})
  (\Delta_1^2+\frac{\Gamma^2}{4})} 
 {\left(\delta (\Delta_1+\Delta_2) +\frac{\kappa'\Gamma}2-
  i\frac{(\Delta_1-\Delta_2)\kappa'}2\right)}
 \langle J_- J_- \rangle\\ 
&+\frac{\Omega_1^*\Omega_2g_a^*g_b}{4(\Delta_2^2+\frac{\Gamma^2}{4})
  (\Delta_1^2+\frac{\Gamma^2}{4})}  {\left(\delta (\Delta_1+\Delta_2)
    +\frac{\kappa'\Gamma}2+
  i\frac{(\Delta_1-\Delta_2)\kappa'}2\right)}
 \langle J_+ J_+ \rangle\biggrb],
\end{aligned}
\end{split}
\label{eq:nab}
\end{equation}
%
%
\begin{equation}
\begin{split}
\frac{d}{dt} \langle J_+J_+\rangle=&-
\frac{\Gamma}{\Delta_1^2+\frac{\Gamma^2}4}
\frac{|\Omega_1|^2}{4}\langle J_+J_+ \rangle
-\frac{\Gamma}{\Delta_2^2+\frac{\Gamma^2}4}
\frac{|\Omega_2|^2}{4}\langle J_+J_+ \rangle  \\
&-\frac{2i N}{\delta^2+\frac{{\kappa'}^2}4}\bigglb[
\begin{aligned}[t]
&\frac{|\Omega_1|^2|g_b|^2}{4(\Delta_1^2+\frac{\Gamma^2}4)^2}
{\left(\Delta_1^2+\frac{\Gamma^2}4\right)}
{\left(\delta+i\frac{\kappa'}2\right)} \langle J_+ J_+
\rangle
+\frac{|\Omega_2|^2|g_a|^2}{4(\Delta_2^2+\frac{\Gamma^2}4)^2}
 {\left(\Delta_2+i\frac{\Gamma}2\right)}^2
 {\left(\delta-i\frac{\kappa'}2\right)}\langle J_+
J_+ \rangle\\ 
&+\frac{\Omega_1\Omega_2^*g_ag_b^*}{4(\Delta_2^2+\frac{\Gamma^2}{4})
  (\Delta_1^2+\frac{\Gamma^2}{4})}
\bigglb(
\begin{aligned}[t]
& {\left(\Delta_1+i\frac{\Gamma}2\right)}
    {\left(\Delta_2-i\frac{\Gamma}2\right)}
    {\left(\delta+i\frac{\kappa'}2\right)} \langle J_- J_+ \rangle\\
&   + {\left(\Delta_1+i\frac{\Gamma}2\right)}
    {\left(\Delta_2+i\frac{\Gamma}2\right)}
    {\left(\delta-i\frac{\kappa'}2\right)} \langle J_+ J_- \rangle
\biggrb)
\biggrb],
\end{aligned}
\end{aligned}
\end{split}
\label{eq:plusplus}
\end{equation}
%
%
and
\begin{equation}
\begin{split}
\frac{d}{dt} \langle J_+J_-\rangle=&
\frac{|\Omega_1|^2}{4(\Delta_1^2+\frac{\Gamma^2}4)}
{\left(\gamma_a\langle \hN_a\rangle-\Gamma \langle J_+J_-
    \rangle\right)}
+\frac{|\Omega_2|^2}{4(\Delta_2^2+\frac{\Gamma^2}4)}
{\left( \gamma_a \langle \hN_b\rangle+\Gamma\langle
    \hN_a\rangle-\Gamma\langle
    J_+J_-\rangle\right)}-\frac{N}{\delta^2+\frac{{\kappa'}^2}4}A\\
\frac{d}{dt} \langle J_-J_+\rangle=&
\frac{|\Omega_1|^2}{4(\Delta_1^2+\frac{\Gamma^2}4)}
{\left(\gamma_b\langle \hN_a\rangle+\Gamma\langle \hN_b\rangle -\Gamma
    \langle J_-J_+   \rangle\right)}
+\frac{|\Omega_2|^2}{4(\Delta_2^2+\frac{\Gamma^2}4)}
{\left( \gamma_b \langle \hN_b\rangle-\Gamma\langle
    J_-J_+\rangle\right)}-\frac{N}{\delta^2+\frac{{\kappa'}^2}4} A,
\end{split}
\label{eq:plusminus}
\end{equation}
where
\begin{equation}
\begin{split}
A=
&\frac{|\Omega_1|^2|g_b|^2}{4(\Delta_1^2+\frac{\Gamma^2}4)^2}
{\left( -\kappa'   \right)}{\left(\Delta_1^2+\frac{\Gamma^2}4   \right)}
\langle J_+ J_- 
\rangle
+\frac{|\Omega_2|^2|g_a|^2}{4(\Delta_2^2+\frac{\Gamma^2}4)^2}
{\left(\Delta_2^2 \kappa' -2\Delta_2\delta\Gamma-\frac{\kappa'\Gamma^2}4  \right)} \langle J_-
J_+ \rangle\\ 
&+\frac{\Omega_1\Omega_2^*g_ag_b^*}{4(\Delta_2^2+\frac{\Gamma^2}{4})
  (\Delta_1^2+\frac{\Gamma^2}{4})} 
 {\left(2i\delta\Delta_1\Delta_2- \delta\Delta_2\Gamma
     +i\Delta_1\frac{\Gamma\kappa'}2-  \frac{\kappa'\Gamma^2}4\right)}  
 \langle J_- J_- \rangle\\ 
&+\frac{\Omega_1^*\Omega_2g_a^*g_b}{4(\Delta_2^2+\frac{\Gamma^2}{4})
  (\Delta_1^2+\frac{\Gamma^2}{4})}  {\left(-2i\delta\Delta_1\Delta_2-
    \delta\Delta_2\Gamma -i\Delta_1\frac{\Gamma\kappa'}2-
    \frac{\kappa'\Gamma^2}4 \right)} 
 \langle J_+ J_+ \rangle.
\label{eq:A}
\end{split}
\end{equation}
\end{widetext}

\end{document}